# Creative Interference and States of Potentiality in Analogy Problem Solving

**Liane Gabora (liane.gabora@ubc.ca) and Adam Saab (adamsaab@gmail.com)**
Department of Psychology, University of British Columbia
3333 University Way, Kelowna BC, V1V 1V7, Canada

## Abstract

Creative processes are widely believed to involve the generation of *multiple, discrete, well-defined* possibilities followed by exploration and selection. An alternative, inspired by parallel distributed processing models of associative memory, is that creativity involves the merging and interference of memory items resulting in a *single* cognitive structure that is *ill-defined*, and can thus be said to exist in a state of potentiality. We tested this hypothesis in an experiment in which participants were interrupted midway through solving an analogy problem and asked what they were thinking in terms of a solution. Naïve judges categorized their responses as AP if there was evidence of merging solution sources from memory resulting in an ill-defined idea, and SM if there was no evidence of this. Data from frequency counts and mean number of SM versus AP judgments supported the hypothesis that midway through creative processing an idea is in a potentiality state.

## Introduction

Creative ideas take time to mature. It is widely assumed that this process involves *searching* through memory and/or *selecting* amongst a set of predefined candidate ideas. For example, computer scientists have modeled the creative process as heuristic search (e.g. Simon, 1973, 1986). In psychology, there is much evidence for, and discussion of, the role of *divergent thinking* in creativity (Guilford, 1968; for a review see Runco, 2010). Divergent thinking is presumed to involve the generation of multiple, often unconventional possibilities. Thus construed, it necessarily goes hand-in-hand with selection, since if you come up with multiple alternatives you eventually weed some of them out. Many well-known theories of creativity, such as the Geneplore model (Finke, Ward, & Smith, 1992), and the Darwinian theory of creativity (Simonton, 1999) involve two stages: generation of possibilities, followed by exploration and ultimately selective retention of the most promising of them. However, the generation stage of creative thinking may be divergent not in the sense that it moves in multiple directions or generates multiple possibilities, but in the sense that it produces a raw idea that is vague or unfocused, that requires further processing to become viable. Similarly, the exploration stage of creative thinking may be *convergent* not in the sense that it entails selecting from amongst alternatives but in the sense that it entails considering a vague idea from different perspectives until it comes into focus. The terms divergent and convergent may be applicable to creative thought not in the sense of going from one to many or from many to one, but in the sense of going from well-defined to ill-defined, and *vice versa*. We hypothesize that although a creative process *may* involve search or selection amongst multiple possibilities, it *need* not, and selection need not figure prominently in a general theory of creativity (Gabora, 2005, 2010).

A central and ubiquitous facet of human creativity is analogy (Gentner, 2010; Mitchell, 1993). Analogies are widely believed to involve two elements: a *source*, which is well understood, and a *target*, which is less well understood but which shares elements with the source. One might suppose that because analogy does not (in general) require that one come up with something new so much as *find a source in memory* that is structurally similar to the target, it is the creative process most likely to involve search or selection. Thus if we can show that even analogy problem solving involves not search or selection amongst predefined alternatives but the resolution of ill-defined states of potentiality, we have fairly strong evidence for the hypothesis that potentiality states figure prominently in the creative process. This paper presents an analogy problem solving experimental test of this.

### Rationale for the Approach

Ontologically, selection amongst multiple well-defined entities entails a different formal structure from actualizing the potential of a single, ill-defined entity (Gabora, 2005; Gabora & Aerts, 2007). Cognitively, thinking of a single vague idea seems relatively straightforward, whereas it is not obvious that one could simultaneously hold in one's mind multiple well-defined ideas. But perhaps the strongest reason to suppose that creativity involves, in the general case, not selection amongst multiple ideas but the honing of a half-baked idea, is that it is consistent with the structure of associative memory (Gabora, 2010). Because memory is sparse, distributed, and content-addressable, knowledge and memories that are relevant to the situation or task at hand naturally come to mind (e.g. Hinton, McClelland, & Rumelhart, 1986; Kanerva, 1988). Neural cell assemblies that respond to the particular features of a situation are activated, and items previously encoded in these cell assemblies (that have similar constellations of features and activate similar distributed sets of neurons), are evoked. Both the vagueness of a 'half-baked' idea and the sense that it holds potential, as well as its capacity to actualize in different ways depending on how one thinks it through, may be side effects of the phenomenon of interference. In interference, a recent memory interferes with the capacity to recall an older memory. A similar phenomenon occurs in neural networks, where it goes by different names: 'crosstalk', 'superposition catastrophe', 'false memories',

'spurious memories', and 'ghosts' (Feldman & Ballard, 1982; Hopfield, 1982; Hopfield, Feinstein, & Palmer, 1983). Interference is generally thought of as detrimental, but it may be of help with respect to creativity. A half-baked idea may be what results when two or more items encoded in overlapping distributions of neural cell assemblies interfere with each other and get evoked simultaneously. We will refer to the phenomenon of interference leading to creative ideation as *creative interference*. When an idea emerges through creative interference, the contributing items are not searched or selected amongst because together they form a single structure. This structure can be said to be in a state of potentiality because its ill-defined elements could take on different values depending on how the analogy unfolds. It is proposed that this unfolding involves disentangling the relevant features from the irrelevant features by observing how the idea looks from sequentially considered perspectives. In other words, one observes how it interacts with various contexts, either internally generated (think it through) or externally generated (try it out).

Theoretical and experimental support for the notion of potentiality states in creative cognition comes from work on concept combination. A mathematical model of concepts that includes potentiality states as a central notion accurately predicts the shifts in applicabilities of properties of concepts when they are placed in different contexts (Aerts & Gabora, 2005a,b). Experiments on concept combination (summarized in Ward & Kolomytz, 2010) show that the more dissimilar the contributing concepts, the more original, yet potentially the less practical, the resulting idea. This suggests that the more ill-defined the unborn idea, the greater the extent to which it exists in a state of potentiality, and the more processing it requires to become viable. Real-time studies of artists and designers indicate that creative ideation involves elaborating on a 'kernel idea' which goes from ill-defined to well-defined through an interaction between artist and artwork (Locher, 2010; Tovey & Porter, 2003; Weisberg, 2004). This too is highly consistent with the notion of potentiality states.

## Theories of Analogy

In tests of analogy solving, the target is presented as a problem that can be solved drawing from the source (Gick & Holyoak, 1983). It is believed that the source tacitly informs the participant, prompting a solution. We now summarize a preeminent theory of how this happens known as *structure mapping*. We then present an alternative that involves the resolution of cognitive states of potentiality.

### Structure Mapping

Structure mapping, in its original formulation, posits that analogy generation occurs in two steps (Gentner, 1983). The first step involves finding an appropriate source in memory and aligning it with the target. The second step involves mapping the correct one-to-one correspondences between the source and the target. Thus structure mapping assumes that once the correct source is found, the analogy making process occurs in isolation from the rest of the contents of the mind. A key principle of structure mapping is the systematicity principle, according to which people prefer to connect structures composed of related predicates.

According to more recent formulation of the theory, the process occurs in three stages (Forbus, Gentner, & Law, 1995; Gentner, 2010). The first stage entails finding all possible source-to-target matches through a quick and dirty process that emphasizes surface similarity. Structural alignment and mapping occur in the second and third stages.

### An 'Actualization of Potential' View of Analogy

While we believe that the basic principles of structure mapping are sensible and well-supported, we call into question some of its underlying assumptions. First, we propose that the first phase is not, as Gentner (2010) puts it, a "structurally blind free-for-all" (p. 753) but rather, constrained by content-addressable structure of associative memory to naturally retrieve items that are in some way (although not necessarily the right or most relevant way) structurally similar (Gabora, 2010). A related assumption we call into question is that alignment and mapping work with discrete, predefined structures. We propose that a 'half-baked' analogy exists in a state of potentiality due to the phenomenon of creative interference. The source may be an amalgam of multiple items that have previously been encoded to the neural cell assemblies activated by the target, and that in the present context cannot be readily separated from one another. The analogy is in a potentiality state because relevant aspects of these items have not yet been disambiguated from aspects that are irrelevant.

### Contrasting the Predictions of Structure Mapping versus Actualizing Potentiality

We have examined two theories concerning the creative process of analogy making. Structure mapping is related to search/selection theories of creativity in emphasizing the challenge of finding a suitable predefined structure. In contrast, according to the 'actualization of potentiality' view, because of the distributed, content-addressable structure of associative memory, suitable pre-defined structures come to mind for free, not as separate and discrete items but merged together, and the challenge is dis-entangling the relevant from the irrelevant.

The two theories give different predictions as to the state of the mind *midway through* analogy formation. This is schematically illustrated in Figure 1. Structure mapping predicts that early on, multiple distinct and separate sources may be identified. Eventually an appropriate source is chosen, but the analogy is still unfinished because not all the correspondences between source and target have been found. There is no reason to expect that the incomplete solution will contain *extra* material.

According to the 'actualization of potentiality' view, however, the incomplete solution is expected to contain *extra* material that would perhaps be correct for similar problems but that is not appropriate for this one. The unfinished solution is ill-defined because irrelevant characteristics of the contributing sources have not yet been disambiguated from characteristics that are relevant.

| | **Structure Mapping** | **Actualization of Potentiality** |
|---|---|---|
| **Early** | Sources Target | Source Target |
| **Late** | Source Target | Source Target |
| **Complete** | Source Target | Source Target |

Figure 1. Highly simplified illustration of the relevant differences between analogy solving by structure mapping versus by actualization of potentiality.

Note that for completed analogies it is not possible to distinguish which theory provides a better explanation of how the analogy was produced. Only for incomplete analogies do the two theories give different predictions.

## Method

We now examine an experimental test of the hypothesis that midway through analogy problem solving the mind is in a state of potentiality. In the following experiment, a procedure that involves stopping participants midway through a problem solving session and asking them what they are thinking (Bower, Farvolden, & Mermigis, 1995), was adapted to analogy problem solving.

### Participants

Eighty-five University of British Columbia undergraduates who were taking first year psychology course participated in the experiment. They received course credit for their participation.

### Materials and Procedures

The source and target for this analogy solving experiment are, respectively, The General, and the Radiation Problem, commonly used one-paragraph-long stories in the analogy literature (Gick & Holyoak, 1983). The General involves a fortress that cannot be captured if all soldiers come from the same direction but that can be successfully captured by dividing the army into small groups of soldiers that converge on the fortress from multiple directions. (The story is provided in Appendix A). The Radiation Problem involves finding a way to destroy a tumor without killing surrounding tissue. (The story is provided in Appendix B). The solution to the Radiation Problem is analogous to the solution to The General; the tumor is destroyed using multiple low-intensity X-rays from different directions.

The experimental procedure consisted of the following:

**Phase One** In the first phase, the *exposure to source phase*, the participants were given five minutes to study The General. They were asked to summarize the story as a test of their story comprehension.

**Phase Two** In the second phase, the *problem solving phase*, the participants were presented with the target, the Radiation Problem. They were given no indication that the story from phase one could help them solve the problem. Since pilot studies showed that the minimum time required to solve the Radiation problem is two minutes, the participants were interrupted after 100 seconds and told they had 20 seconds to write down whatever they were currently thinking in terms of a solution.

**Followup** In a questionnaire distributed immediately afterward participants were asked whether they noticed a relation between The General and The Radiation Problem, and if so, at what point they noticed it.

### Judging

Results of both phases were assessed by six judges who were naïve as to the theoretical rationale for the experiment. The story summaries produced in phase one were judged for comprehension on a three-point scale: poor, fair, or good.

Since we were interested in the nature of cognitive states midway through a creative process, participants who correctly solved the problem in the allotted time were removed from the analysis ($N$ = 34). They were deemed to have correctly solved the problem if they found all three of the correspondences provided in Table 1.

Table 1: The necessary correspondences for a complete analogical solution.

| The General (Source) | Radiation Problem (Target) |
|---|---|
| 1. Multiple groups of soldiers | 1. Multiple rays |
| 2. Small groups | 2. Low intensity |
| 3. Groups converge from different directions | 3. Rays converge from different directions |

The judges were asked to categorize each of the remaining incomplete solutions as either Structure Mapping (SM) or Actualization of Potentiality (AP) according to the characteristics of each provided in Table 2.

Table 2: Characteristics used to judge incomplete analogy solutions as Structure Mapping (SM) versus Actualization of Potentiality (AP).

| Structure Mapping (SM) | Actualization of Potentiality (AP) |
|---|---|
| 1. If multiple solutions are given they are considered separate and distinct (for example, separated by the word 'or') | 1. If multiple solutions are given they are jumbled together |
| 2. Does not contain extra, irrelevant information | 2. Contains extra information that would be relevant for related problems but that is not relevant for this one |

A potential concern at this point is that an answer might contain extraneous information because it was elaborated following retrieval, rather than because of creative interference. There is evidence that analogy making does in some cases involve adapting or elaborating the source to improve the match (e.g. Ross, 1987). Structure mapping does not emphasize this kind of adaptation or elaboration, but it allows for and is not incompatible with it. However, in the analogy used here, no adapting or elaborating of the source was needed to generate a complete and correct solution. In other words, if the correct source (The General story) was found, it could be used as is, without elaboration. Therefore, if extraneous information is present we have good reason to believe that it was due to creative interference.

An example of an answer that was categorized as structure mapping (SM) is:

> "No idea. Don't know much about science. Maybe try to have a low-intensity ray that would sufficiently kill the tumor but not destroy healthy tissues."

In this answer, one of correspondences has been found (correspondence 2: low intensity ray). Since the other two correspondences were not found (multiple rays and different directions) the solution is incomplete. Since the answer provides no evidence that the participant's current conception of a solution consists of multiple items jumbled together in memory, it was classified as SM.

We now present two examples of answers that were categorized as AP. The first example is:

> "First, what kind of tissue will be destroyed with the ray treatment? Can it be replaced using skin graft? How much tissue will be destroyed in the surrounding area? Will the cost outweigh the benefit? This needs to be considered if using the full strength ray."

In this incomplete solution, none of the three correspondences has been found. However, the phrase "if using the full strength ray" indicates that the participant has considered, or is about to consider, the possibility of using a ray that is less than full strength, which suggests that correspondence 2 is within reach. It was classified as AP because it includes irrelevant information (such as concern about the kind of tissue) activated by the target that is unnecessary to generation of the correct solution.

A second example of an answer that was categorized as AP is the following:

> "The high intensity ray is necessary to kill the tumor so maybe shooting it in short successive bursts from different angles will kill the tumor without killing too much healthy tissue."

In this incomplete solution, one of the correspondences has been found: different directions (correspondence 3). It was classified as AP because it also includes irrelevant information (the notion of "short successive bursts") activated by the target that is unnecessary to generation of the correct solution.

## Results

The intraclass correlation for the degree of agreement concerning ratings of story comprehension across the judges was .704, which indicates that they are reliable. 36 participants responded that they saw a relation between the story and the problem, and 15 said they did not. A Kruskal-Wallis test was conducted to determine whether there was an interaction between the poor, fair, and good ratings of story comprehension, and which theory was supported by a particular incomplete analogy. The test was not significant, $\chi^2(2, N = 50) = 3.65$, $p = .16$, $r = .27$. This indicates that there was no interaction between story comprehension and which theory was supported by a response.

Each of the 51 incomplete solutions (those that remained after complete solutions were removed from the analysis) was classified as supportive of structure mapping if 4 or more judges judged it as structure mapping, and as supportive of actualizing potentiality if 4 or more judges judged it as actualizing potentiality. In cases where judging was tied (N = 8) random number generation was used to assign case values. As shown in Figure 2, 39 were classified as supportive of actualizing potentiality, and 12 were classified as supportive of structure mapping. A one-sample chi-square test was significant, $\chi^2(1, N = 50) = 1.43$, $p < .001$. Thus the frequency count data support the hypothesis that analogies are generated by actualizing potentiality.

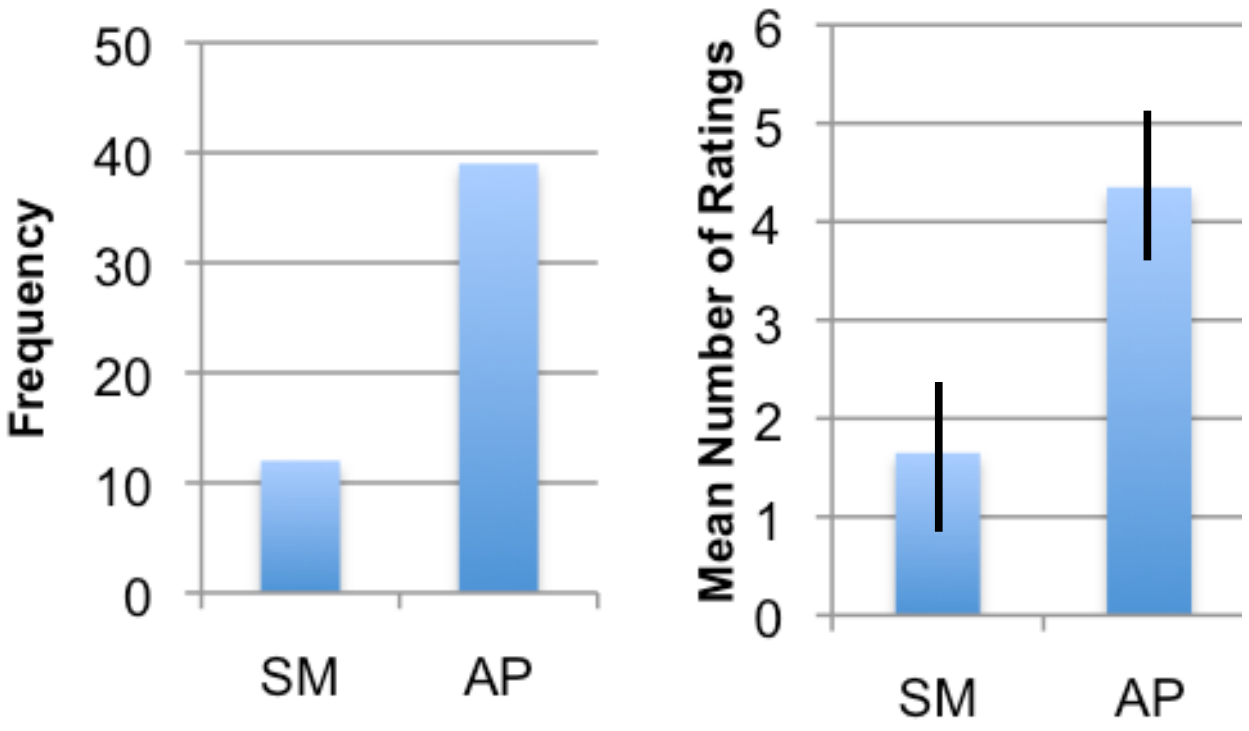

Figure 2: (a) Frequency count of solution judgments for structure mapping (SM), on the left ($N = 12$) and actualization of potentiality (AP), on the right ($N = 39$). (b) Mean number of ratings of SM, on the left (M = 1.65, SD =1.66) and AP, on the right (M = 4.35, SD =1.66).

A further analysis compared the mean number of judgments (out of a maximum of 6, the total number of judges) across all responses that supported each theory. Taking the mean across all 51 responses, the mean number of structure mapping judgments was 1.65 (*SD* =1.66), and the mean number of actualizing potentiality judgments was 4.35 (*SD* =1.66). A paired-sample $t$-test showed that the difference was significant $t(50) = -5.82$, $p < .001$), and the effect size ($\eta^2 = .25$) was large. Thus these data corroborate the above frequency count findings. The mean judgment scores for structure mapping and actualizing potentiality are given in Figure 2b.

## Discussion

Although framed as an alternative to structure mapping, the data presented here are not incompatible with its key principles of structural alignment, systematicity, and mapping. However, we propose that the sparse, distributed, content-addressable structure of associative memory ensures that any item that comes to mind as a potential source bears some structural similarity (deep or superficial) to the target. Thus the initial stage of analogy solving is viewed as not structurally blind, but confused due to the multitude of potentially relevant structures that present themselves. Due to the phenomenon of creative interference, what comes to mind may be quite unlike anything that has ever been experienced. Both the data from frequency counts and mean number of SM versus AP judgments support the prediction that a half-baked analogy exists in a state of potentiality, in which the constituent items are merged together, as opposed to a collection of candidate items each in the separate and distinct form in which they were originally encoded in memory. The vagueness or 'half-baked' quality reflects that it is still uncertain how, in the context of each other, these elements come together as a realizable whole.

A limitation of the current study is that it only involved one analogy-solving task. Future studies will investigate the role of potentiality states in other analogy solving tasks, as well as other kinds of creative processes, and in human cognition more generally.

## Acknowledgments

We thank Brian O'Connor and Tomas Veloz for comments. We are grateful for funding to Liane Gabora from the *Social Sciences and Humanities Research Council of Canada* and the *Fund for Scientific Research,* Belgium.

## References

Aerts, D. & Gabora, L. (2005). A state-context-property model of concepts and their combinations I: The structure of the sets of contexts and properties. *Kybernetes, 34*(1&2), 167-191.

Aerts, D. & Gabora, L. (2005). A state-context-property model of concepts and their combinations II: A Hilbert space representation. *Kybernetes, 34*(1&2), 192-221.

Bowers, K. S., P. Farvolden, & L. Mermigis. (1995). Intuitive Antecedents of Insight. In S. M. Smith, T. B. Ward, & R. A. Finke (Eds.), *The creative cognition approach* (pp. 27-52). Cambridge MA: MIT Press.

Feldman, J., & Ballard, D. (1982). Connectionist models and their properties. *Cognitive Science, 6,* 204–254.

Finke, R. A., Ward, T. B., & Smith, S. M. (1992). *Creative cognition: Theory, research, and applications*. Cambridge, MA: MIT Press.

Forbus, K. Gentner, D. & Law, K. (1995). MAC/FAC: A model of similarity-based retrieval. *Cognitive Science, 19,* 141-205.

Gabora, L. (2005). Creative thought as a non-Darwinian evolutionary process. *Journal of Creative Behavior, 39*(4), 262-283.

Gabora, L. & Aerts, D. (2007). A cross-disciplinary framework for the description of contextually mediated change. *Electronic Journal of Theoretical Physics, 4*(15), 1-22.

Gabora, L. (2010). Revenge of the 'neurds': Characterizing creative thought in terms of the structure and dynamics of human memory. *Creativity Research Journal, 22*(1), 1-13.

Gentner, D. (1983). Structure-mapping: A theoretical framework for analogy. *Cognitive Science*, *7(2)*, 155-170.

Gentner, D. (2003). Why We're So Smart. In D. Gentner and S. Goldin-Meadow (Eds.), *Language in mind: advances in the study of language and thought* (pp. 195-236). Cambridge MA: MIT Press.

Gentner, D. (2010). Bootstrapping the mind: Analogical processes and symbol systems. *Cognitive Science, 34,* 752-775.

Gick, M. L. & Holyoak, K. J. (1983). Schema induction and analogical transfer. *Cognitive Psychology*, *15(1)*, 1-38.

Guilford, J. P. (1968). *Intelligence, creativity and their educational implications*. San Diego: Knapp.

Hinton, G., McClelland, J., & Rumelhart, D. (1986). Distributed representations. In D. Rumelhart, J. McClelland, & the PDP research Group (Eds.), *Parallel distributed processing, Volume 1* (pp. 77-109). Cambridge, MA: MIT Press.

Hopfield, J. J. (1982). Neural networks and physical systems with emergent collective computational abilities. *Proceedings of the National Academy of Sciences (Biophysics), 79,* 2554–2558.

Hopfield, J. J., Feinstein, D. I., & Palmer, R. D. (1983). ''Unlearning'' has a stabilizing effect in collective memories. *Nature, 304,* 159–160.

Kanerva, P. (1988). *Sparse, distributed memory*. Cambridge, MA: MIT Press.

Locher, P. (2010). How does a visual artist create an artwork? In (J. Kaufman & R. Sternberg, Eds.) *Cambridge Handbook of Creativity*. (pp. 131-144). Cambridge UK: Cambridge University Press.

Mitchell, M. (1993). *Analogy-making as perception: A computer model.* Cambridge, MA: MIT Press.

Ross, B. H. (1987). This is like that: The use of earlier problems and the separation of similarity effects. *Journal of Experimental Psychology: Learning, Memory and Cognition, 13,* 629-639.

Runco, M. (2010). Divergent thinking, creativity, and ideation. In (J. Kaufman & R. Sternberg, Eds.) *Cambridge Handbook of Creativity*. (pp. 413-446). Cambridge UK: Cambridge University Press.

Simon, H. A. (1973). Does scientific discovery have a logic? *Philosophy of Science, 40*, 471–480.

Simon, H. A. (1986). Understanding the processes of science: The psychology of scientific discovery. In T. Gamelius (Ed.), *Progress in science and its social conditions* (pp. 159–170). Oxford: Pergamon Press.

Simonton, D. K. (1999). Creativity as blind variation and selective retention: Is the creative process Darwinian? *Psychological Inquiry*, *10*, 309–328.

Tovey, M. & Porter, S. (2003). Sketching, concept development, and automotive design. *Designs Studies, 24,* 135-153.

Ward, T. & Kolomytz, Y. (2010). Cognition and creativity. In (J. Kaufman & R. Sternberg, Eds.) *Cambridge Handbook of Creativity*. (pp. 93-112). Cambridge UK: Cambridge University Press.

Weisberg, R. (2004). Structure in the creative process: A quantitative case-study of the creation of Picasso's Guernica. *Empirical Studies of the Arts, 22,* 23-54.

## Appendix A: The General

The following story, referred to as The General, was used as the source in this analogy solving experiment:

A small country was ruled from a strong fortress by a dictator. The fortress was situated in the middle of the country, surrounded by farms and villages. Many roads led to the fortress through the countryside. A rebel general vowed to capture the fortress. The general knew that an attack by his entire army would capture the fortress. He gathered his army at the head of one of the roads, ready to launch a full-scale direct attack. However, the general then learned that the dictator had planted mines on each of the roads. The mines were set so that small bodies of men could pass over them safely, since the dictator needed to move his troops and workers to and from the fortress. However, any large force would detonate the mines. Not only would this blow up the road, but it would also destroy many neighboring villages. It therefore seemed impossible to capture the fortress. However, the general devised a simple plan. He divided his army into small groups and dispatched each group to the head of a different road. When all was ready he gave the signal and each group marched down a different road. Each group continued down its road to the fortress so that the entire army arrived together at the fortress at the same time. In this way, the general captured the fortress and overthrew the dictator.

## Appendix B: The Radiation Problem

The following story, referred to as The Radiation Problem, was used as the target in this experiment.

Suppose you are a doctor faced with a patient who has a malignant tumor in his stomach. It is impossible to operate on the patient, but unless the tumor is destroyed the patient will die. There is a kind of ray that can be used to destroy the tumor. If the rays reach the tumor all at once at a sufficiently high intensity, the tumor will be destroyed. Unfortunately, at this intensity the healthy tissue that the rays pass through on the way to the tumor will also be destroyed. At lower intensities the rays are harmless to healthy tissue, but they will not affect the tumor either. What type of procedure might be used to destroy the tumor with the rays, and at the same time avoid destroying the healthy tissue?